\documentclass[
 aip,
 amsmath,amssymb,
reprint,
]{revtex4-1}

\usepackage{graphicx}
\usepackage{dcolumn}
\usepackage{bm}

\usepackage[utf8]{inputenc}
\usepackage[T1]{fontenc}
\usepackage{mathptmx}
\usepackage{siunitx}
\begin{document}

\preprint{AIP/123-QED}

\title[]{Coherence and indistinguishability of highly pure single photons from non-resonantly and resonantly excited telecom C-band quantum dots}

\author{C. Nawrath}
\email{c.nawrath@ihfg.uni-stuttgart.de}
\homepage{www.ihfg.uni-stuttgart.de}\noaffiliation
\author{F. Olbrich}\noaffiliation
\author{M. Paul}\noaffiliation
\author{S. L. Portalupi}\noaffiliation
\author{M. Jetter}\noaffiliation
\author{P. Michler}\noaffiliation
\affiliation{Institut f\"ur Halbleiteroptik und Funktionelle Grenzfl\"achen, \\Center for Integrated Quantum Science and Technology (IQ\textsuperscript{ST}) and SCoPE, \\University of Stuttgart, Allmandring 3, 70569 Stuttgart, Germany}

\date{\today}

\begin{abstract}
In the present work, the effect of resonant pumping schemes in improving the photon coherence is investigated on InAs/InGaAs/GaAs quantum dots emitting in the telecom C-band. The linewidths of transitions of multiple exemplary quantum dots are determined under above-band pumping and resonance fluorescence via Fourier-transform spectroscopy and resonance scans, respectively. The average linewidth is reduced from \SI{9.74}{\giga \hertz} in above-band excitation to \SI{3.50}{\giga \hertz} in resonance fluorescence underlining its superior coherence properties. Furthermore, the feasibility of coherent state preparation with a fidelity of \SI{49.2}{\percent} is demonstrated, constituting a step towards on-demand generation of coherent, single C-band photons from quantum dots. Finally, two-photon excitation of the biexciton is investigated as a resonant pumping scheme. A deconvoluted single-photon purity value of $g^{(2)}_{\mathrm{HBT}}(0)=0.072\pm 0.104$ and a degree of indistinguishability of $V_{\mathrm{HOM}}=0.894\pm0.109$ are determined for the biexciton transition. This represents an important step towards fulfilling the prerequisites for quantum communication applications like quantum repeater schemes at telecom wavelength.
\end{abstract}

\maketitle

Over the past two decades, semiconductor quantum dots (QDs) have received unceasing attention from researchers in the field of quantum optics due to their outstanding properties in terms of non-classical light emission~\cite{Ding,Somaschi,Dous,Huber2017,Muller}, i.e. bright single-photon emission, entanglement fidelity, indistinguishability and the simultaneous combination of the aforementioned~\cite{Muller,Huber2017}. This designates them as promising candidates for applications like quantum computing and quantum communication~\cite{QCom}. The best performances are currently achieved with GaAs-based dots emitting in the near infrared~\cite{Michler2017}. However, especially regarding quantum communication schemes, an emission wavelength around \SI{1550}{\nano \meter} (Telecom C-band) is much sought-after both for satellite-based quantum communication due to an atmospheric transmission window and the possibility to perform it in broad daylight~\cite{Liao2017}, as well as for its fiber-based counterpart due to the global absorption minimum and low dispersion of standard glass fibers forming the existing global fiber network~\cite{Agra}. However, to extend the range of quantum communication applications such as quantum key distribution~\cite{Scar,Takemoto2010}, quantum relays~\cite{Jacobs2002,Huwer2017} or quantum repeaters~\cite{Briegel1998,Jones,Duan2001a,Sangouard2011} are needed. The ideal light source for such applications combines bright single-photon and entangled-photon pair emission with a high degree of indistinguishability at \SI{1550}{\nano \meter}.

The emission of single and entangled photons in the telecom C-band has been demonstrated in two material systems, namely InAs/InP~\cite{Muller2018,Benyoucef2013} and InAs/InGaAs/GaAs~\cite{Paul2017,Olbr}. The last requirement, i.e. the indistinguishability of photons, is of major importance because it is necessary for two-photon interference (TPI), enabling linear-optic Bell state measurements and, therefore, entanglement swapping~\cite{Zukowski1993,Pan} in quantum repeater schemes. An experimental demonstration at this wavelength has been elusive in both material systems up to now.\\
However, long coherence times and the teleportation of a quantum state have been demonstrated in the InAs/InP system~\cite{Anderson2019}, promising a high degree of indistinguishability. For QDs based on GaAs on the other hand, a straightforeward implementation of distributed Bragg reflectors (DBRs) offers the prospect of fabricating high-quality cavities and micro-pillars with a high extraction efficiency~\cite{Dous,Somaschi,Ding,Unsleber2016}. Furthermore, the recently demonstrated feasibility of strain-tuning~\cite{Zeuner2018} paves the way to tune different QDs into resonance, facilitating remote TPI experiments~\cite{Patel2010,Weber2018}. For practical applications, the coherence as a major impact on the indistinguishability of the emitted photons is of crucial importance, not least since the latter in turn limits the TPI visibility. Apart from properties inherent to the sample structure like the presence of charge carrier trap states~\cite{Kamada2008,Houl}, the coherence and indistinguishability are strongly influenced by the optical pumping scheme~\cite{Ates2009,Kalliakos2016}. Among the possible schemes, resonant ones such as resonance fluorescence (RF)~\cite{Muller2007} and two-photon excitation (TPE)~\cite{Brunner1994,Stufler2006,Jaya,Muller} are known to be most favorable for the optical properties of the emission. In the latter, the biexciton (XX) is directly pumped via two-photon absorption over a virtual state and can decay back to the ground state via the exciton (X) as an intermediate step. Since this cascade can result in the emission of polarization-entangled photon pairs~\cite{Bens} and the XX is resonantly excited, TPE can simultaneously yield excellent results in terms of single-photon purity, entanglement fidelity and indistinguishability~\cite{Muller,Huber2017}.

To quantify the advantages of resonant excitation, a study on the coherence properties under three different excitation schemes is performed. The charge carriers are pumped either above the band gap of the barrier material (above-band, AB), or in RF or via TPE. When pumping in AB and RF, the linewidth is investigated by means of Fourier-transform spectroscopy and RF scans, respectively. Under TPE, on the other hand, the single-photon purity and the degree of indistinguishability, which in turn is strongly impacted by the coherence of the photons, is determined. On top of these measurements performed in continuous-wave (cw) excitation, pulsed RF is performed to coherently prepare the excited state and investigate the state preparation fidelity. For all measurements, the sample is mounted in a He flow cryostat, cooled to \SI{4}{\kelvin} and optically excited with a conventional confocal microscopy setup. For the Fourier-transform spectroscopy measurements, a Michelson interferometer (MI) with variable delay length of up to $\pm$\SI{75}{\milli \meter} is used.  In RF and TPE, the microscope setup is used in dark-field mode, filtering out the laser light based on its polarization~\cite{Kuhlmann2013}. For the measurements on the indistinguishability, an unbalanced, fiber-based Mach-Zehnder interferometer (MZI) is used. The sample under investigation is based on GaAs and employs a metamorphic buffer layer of InGaAs with a gradually increasing In-content to shift the emission of the InAs QDs to the telecom C-band. The capping layer consists of InGaAs. Furthermore, 20 distributed Bragg reflector pairs, consisting of AlAs/GaAs, are used to enhance the brightness of the sample. More details on the structure and growth conditions can be found in Ref.~\onlinecite{Paul2017}.

\begin{figure}
\includegraphics{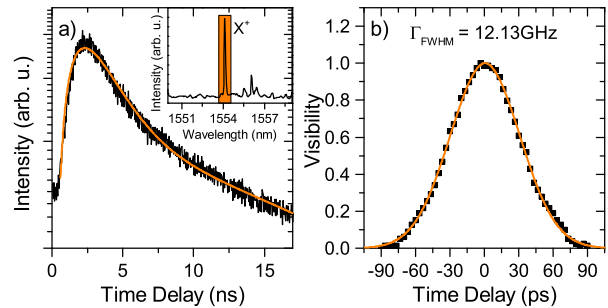}% Here is how to import EPS art
\caption{\label{fig:above_band} Above-band pumping: a) TCSPC measurement on an exemplary QD yielding a fast decay time of \SI{1.56}{\nano \second}. The corresponding spectrum is shown in the inset where the orange area indicates the width of the spectral transmission window of the monochromator used for the experiment. b) Visibility of the interference fringes of a Michelson interferometer over the temporal path length difference alongside a Voigt fit. The given linewidth stands for the FWHM of the Voigt profile. The homogeneous (inhomogeneous) contribution amounts to \SI{0.30}{\giga \hertz} (\SI{11.97}{\giga \hertz}) for this particular QD transition.}
\end{figure}

The coherence time $T_{2}$ and the linewidth $\Gamma_{\mathrm{FWHM}}$, taken as the full width at half maximum (FWHM), depend on the radiative lifetime $T_{1}$ of the excitonic state and the dephasing time $T_{2}^{*}$ via~\cite{Michler2017} $\Gamma_{\mathrm{FWHM}}\propto1/T_{2}=1/(2T_{1})+1/T_{2}^{*}$. If only the homogeneous broadening due to the limited radiative lifetime is present, i.e. $T_{2}=2T_{1}$, one speaks of Fourier transform-limited (FT) emission resulting in a Lorentzian lineshape. The dephasing time $T_{2}^{*}$ includes further homogeneous broadening effects due to interactions with the phonon bath, which increase the linewidth of the Lorentzian, as well as inhomogeneous broadening effects like an instable electrical and magnetic environment~\cite{Kuhlmann2013a} of the QD leading to a Gaussian contribution to the lineshape. The case that both broadening types are present results in a Voigt profile allowing, for sufficient spectral resolution, to access the contributions of both types of broadening to the total lineshape.\\
Firstly, the decay dynamics are investigated using time-correlated single photon counting (TCSPC) measurements in AB excitations on 12 representative QD transitions, which are predominantly positively charged excitons (X$^{+}$) for this sample~\cite{Carmesin2018}. A similar dynamic behaviour, exemplarily displayed in figure \ref{fig:above_band}a), is observed for all QD, i.e. a rise time on the order of \SI{1}{\nano \second}, followed by a fast exponential decay with an average time constant of \SI{1.71}{\nano \second}. Three quarters of the investigated dots exhibit a secondary exponential decay with a mean time constant of \SI{8.94}{\nano \second} and a standard deviation of \SI{3.6}{\nano \second}. The contribution of the primary decay to the overall signal is between half an order and three orders of magnitude stronger than its secondary counterpart.\\
This strong variation and the large standard deviation of the secondary time constant, point to a local effect like the presence of charge carrier trap states refilling the QD~\cite{Carmesin2018}, as an explanation for the slow decay. The possible presence of non-radiative decay channels is assumed to be connected to local effects, as well, which would be reflected in a large spread of the values determined for the fast decay constant from different QDs. The standard deviaton, however, only yields a value of \SI{0.46}{\nano \second}, justifying the assumption that non-radiative recombination channels can be neglected. Because the measurements were performed close to saturation, double excitations and state filling effects~\cite{Munoz_Matutano_2014} can explain the slow rise time observed in most measurements. Individual QDs, however, exhibit a significantly shorter rise time. For this reason, the observation of a slow rise time is ascribed to the experimental conditions, rather than intrinsic effects. In particular the intradot relaxation to the s-shell is assumed to be fast, which is typical for In(Ga)As QDs~\cite{Kurtze2009}. Under these assumptions, the fast decay time can be used as an estimation of the radiative lifetime time $T_{1}$, which yields Fourier-limited values of the coherence time $T_{2,\mathrm{FT}}=\SI{3.42\pm0.92}{\nano \second}$ and the linewidth $\Gamma_{\mathrm{FWHM,FT}}=1/(\pi T_{2,\mathrm{FT}})=\SI{0.1\pm0.03}{\giga \hertz}$. The errors refer to one standard deviation $\sigma$ and the same conventions as in the supplementary of Ref.~\onlinecite{Versteegh2014a} are used. An overview over all coherence properties determined throughout this work is given in table \ref{tab:coherence}.

The linewidth $\Gamma_{\mathrm{FWHM}}$ in AB pumping is evaluated via Fourier-transform spectroscopy using a MI on 9 QDs. The result of this measurement on the same QD as shown in figure \ref{fig:above_band}a) is depicted in figure \ref{fig:above_band}b). When fitting the visibility of the interference fringes over the temporal delay, i.e. the first-order coherence function $g^{(1)}(\tau)$, with the Fourier transform of a Voigt profile (orange), both, the overall linewidth $\Gamma_{\mathrm{FWHM}}$ as well as the contributions $\Gamma_{\mathrm{hom}}$ ($\Gamma_{\mathrm{inhom}}$) due to homogeneous (inhomogeneous) broadening, can be evaluated. The mean value of the overall linewidth is \SI{9.74}{\giga \hertz}, with a standard deviation of \SI{3.29}{\giga \hertz}. The coherence time $T_{2}$ can be calculated via~\cite{Loudon:83} $T_{2}=\int_{-\infty}^{\infty}|g^{(1)}(\tau)|^{2}\mathrm{d}\tau$ (see table \ref{tab:coherence}). The mean homogeneous linewidth $\Gamma_{\mathrm{hom}}$ is \SI{0.98\pm 0.82}{\giga \hertz}. The discrepancy between this value and $\Gamma_{\mathrm{FWHM,FT}}$, as calculated from the TCSPC measurements, is due to homogeneous broadening mechanisms other than the finite radiative decay time. As expected, the inhomogeneous broadening mechanisms are the dominant source of decoherence in AB pumping. This can presumably be attributed to the noisy electrical environment created by the optically excited charge carriers in the barrier material and wetting layer and the phonon-assisted relaxation processes from other QD states prior to emission.
\begin{table}
\caption{\label{tab:coherence}Overview over the coherence properties: linewidth values determined in AB excitation via Fourier-transform spectroscopy and in RF via resonance scans. Radiative lifetime and coherence properties for Fourier-limited (FT) emission determined via AB TCSPC measurements. In all cases, the average values ($\varnothing$), the standard deviation $\sigma$ and the best value, as the most coherent one measured, are given.}
\begin{ruledtabular}
\begin{tabular}{c|ccc|ccc}
Scheme & \multicolumn{3}{c|}{AB} & \multicolumn{3}{c}{RF}\\\hline
Value& $\varnothing$ & $\sigma$ & best value &$\varnothing$ & $\sigma$ & best value\\\hline
$\Gamma_{\mathrm{FWHM}}$~(GHz) & 9.74 & 3.29 & 4.47 & 3.50 & 0.39 & 2.78 \\
$T_{2}$~(ns)& 0.073&0.030&0.144&0.176&0.025&0.220\\
$\Gamma_{\mathrm{inhom}}$~(GHz) & 9.31 & 3.43 & 4.37 & 3.28 & 0.33 & 2.63 \\
$\Gamma_{\mathrm{hom}}$~(GHz)& 0.98 & 0.82 & 0.28 & 0.40 & 0.21 & 0.16 \\\hline
$T_{1}$~(ns)& 1.71 & 0.46 & & & & \\
$\Gamma_{\mathrm{FWHM,FT}}$~(GHz) & 0.1 & 0.03 & & & &\\
$T_{2,\mathrm{FT}}$~(ns)& 3.42 & 0.92 &  & & &\\
\end{tabular}
\end{ruledtabular}
\end{table}

\begin{figure}
	\includegraphics{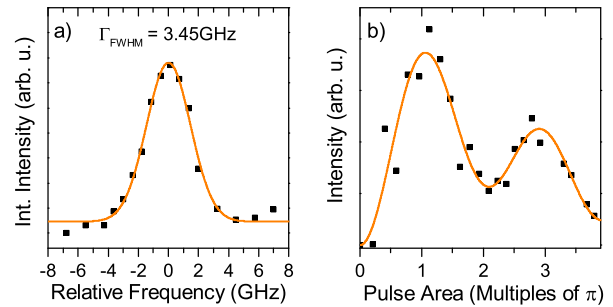}% Here is how to import EPS art
	\caption{\label{fig:rf} Resonance fluorescence: a) Scan of the excitation laser frequency over an exemplary QD transition in natural frequency relative to the maximum at \SI{1548.01}{\nano \meter}. The data is fitted with a Voigt profile (orange). The homogeneous (inhomogeneous) contributions yield \SI{0.78}{\giga \hertz} (\SI{3.49}{\giga \hertz}) for this particular transition. b) Rabi oscillations measured in pulsed RF.}
\end{figure}

Since in RF the charge carriers are directly excited to the discrete QD states, both of these processes are circumvented. To evaluate the linewidth, the excitation frequency is scanned over the QD transition and the integrated intensity is recorded. The laser linewidth of \SI{40}{\mega \hertz} is small enough to forgo a deconvolution. An exemplary scan is depicted for one out of five investigated QDs in figure \ref{fig:rf}a) in natural frequency relative to the resonance of \SI{1548.01}{\nano \meter}. The data from all scans are fitted with a Voigt profile. The mean linewidth amounts to \SI{3.50}{\giga \hertz}, with a standard deviation of \SI{0.39}{\giga \hertz}. The homogeneous linewidth contribution yields an average of \SI{0.4}{\giga \hertz} and is larger than the Fourier-limited linewidth determined via TCSPC measurements. As in AB excitation, this is due to homogeneous broadening mechanisms other than the finite radiative decay time. As expected, in RF a considerable improvement of the coherence time is observed. The remaining inhomogeneous linewidth could be due to an instable magnetic field caused by randomly oscillating spins~\cite{Kuhlmann2013a} and due to a random occupation and depletion of charge carrier trap states by background charges~\cite{Kamada2008,Houl}. Investigations on the temperature dependence of the emission~\cite{Carmesin2018} suggest the presence of such trap states in close spatial proximity to some QDs of this sample.\\
Moreover, the feasibility of coherent state preparation is proven by Rabi oscillations visible in the plot of the integrated intensity over the pulse area in figure \ref{fig:rf}b). To fit the data, the optical Bloch equations are solved numerically with an additional decay channel~\cite{Wang:05}. From this, the state preparation fidelity of this process is determined to \SI{49.2}{\percent}. This paves the way to on-demand generation of single, coherent QD C-band photons.

\begin{figure}
	\includegraphics{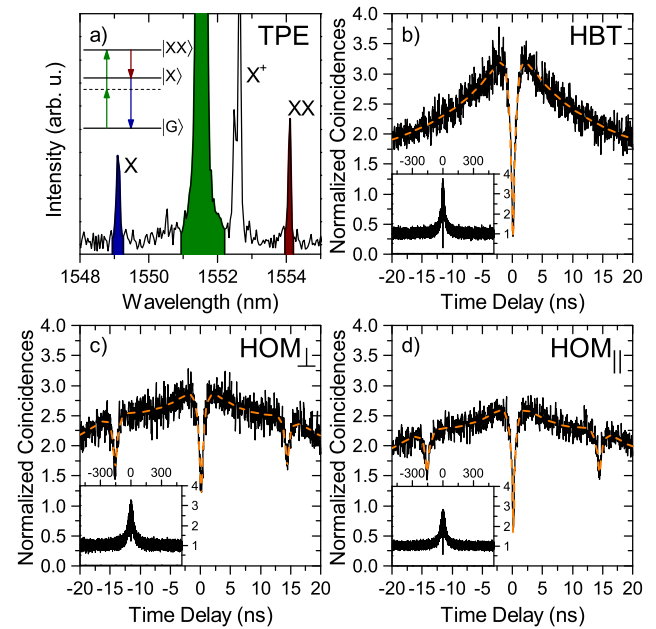}% Here is how to import EPS art
	\caption{\label{fig:tpe} Two-photon excitation: a) Spectrum and energy diagram showing TPE. b) Second order intensity autocorrelation measurement with fit funcion (orange). c) and d) TPI of distinguishable and indistinguishable photons in TPE with the respective fit functions (orange). The insets of b)-d) show the same data with a correlation window of $\pm$\SI{500}{ns}.}
\end{figure}

Combining the advantage of resonant state preparation and the radiative decay via the XX-X cascade, TPE has been identified as a promising form of excitation~\cite{Muller,Huber2017}. The energy scheme and the corresponding spectrum are displayed in figure \ref{fig:tpe}a). One can clearly see the laser (green) and, in symmetric energetic distance to it, the peaks from the X and XX displayed in blue and dark red. The matching integrated intensity of the XX and the X line is a footprint of TPE. The spectral feature around \SI{1552.5}{\nano \meter} stems partly from the scattered laser and partly from the X$^{+}$ that is pumped via the phonon sideband. The following measurements are performed on the XX line.
\\
In figure \ref{fig:tpe}b) a Hanbury-Brown and Twiss (HBT) measurement of the second-order correlation function $g^{(2)}_{\mathrm{HBT}}(\tau)$ is shown. Superimposed on the expected antibunching dip at zero time delay, a strong bunching can be observed and needs to be taken into account when normalizing the data to the Poissonian level (see insets in figure \ref{fig:tpe} for long time delays). The best agreement between the data and a fit function is achieved when including three distinct processes leading to bunching. The fit function applied here reads~\cite{Sallen2010}

\begin{eqnarray}
g^{(2)}_{\mathrm{HBT}}(\tau)=&&a\left(1-b\cdot \exp\left(-\frac{\left|\tau-\tau_{0} \right|}{T_{b}}\right)\right)\nonumber\\
&&\times \prod_{i=1}^{3}\left(1+c_{i}\cdot \exp\left(-\frac{\left|\tau-\tau_{0} \right|}{T_{c,i}}\right)\right),
\label{eq:hbt}
\end{eqnarray}

with $a,b,c_{i},T_{c,i}$, and $\tau_{0}$ as fitting parameters. The parameter $T_{b}$ depends on the radiative lifetime and the pumping rate. The resulting time constants for the bunching are $T_{c,1}=\SI{6.63\pm2.0}{\nano \second}$, $T_{c,2}=\SI{23.99\pm2.2}{\nano \second}$ and $T_{c,3}=\SI{116.8\pm26.5}{\nano \second}$. Possible reasons for this behaviour are spectral diffusion due to background charge carriers, phonon-assisted laser re-excitation of the XX from the X, spin flips rendering a bright X in a dark state and vice versa, fluctuations of the local magnetic field due to nuclear spins and/or background carriers randomly occupying the QD states, impeding the excitation of the XX. Bunching due to blinking is usually observed in RF~\cite{Weber2018}. The fit according to equation \ref{eq:hbt} yields $g^{(2)}_{\mathrm{HBT,raw}}(0)=0.102\pm0.109$ as a raw value. When the data are deconvoluted with the Gaussian-shaped system response function of the detectors and the electronics ($\mathrm{FWHM}=\SI{93}{\pico \second}$ measured via the autocorrelation of a picosecond laser pulse), a value of $g^{(2)}_{\mathrm{HBT,decon}}(0)=0.072\pm0.104$ is obtained, confirming the high single-photon purity expected for TPE of a QD. The errors are calculated via error propagation from the $1\sigma$-confidence bounds of the fitting parameters determined by the non-linear least squares fitting algorithm.\\
To evaluate the indistinguishability of the emitted XX photons in TPE, an unbalanced, fiber-based Mach-Zehnder interferometer with a delay line of \SI{14.3}{\nano \second} is used. To resolve the degree of the Hong-Ou-Mandel (HOM) effect~\cite{Hong1987} expected for indistinguishable photons, the contrast between the autocorrelation measurement with co-polarized (indistinguishable) and cross-polarized (distinguishable) photons, is evaluated. For this, the polarization of the photons can be changed independently in the two interferometer arms. The HOM visibility, i.e. the degree of indistinguishability, is then calculated as $V_{\mathrm{HOM}}=1-g^{(2)}_{\parallel}(0)/g^{(2)}_{\perp}(0)$ from the zero-delay autocorrelation values of the indistinguishable and distinguishable case. To fit the data, the conventional equation for HOM measurements in continuous wave excitation~\cite{Patel2008} is used, inserting however equation \ref{eq:hbt} for $g^{(2)}_{\mathrm{HBT}}(\tau)$ to account for the bunching behaviour. As expected, the obtained bunching time scales are similar to the ones in the HBT measurement. Since the bunching behaviour differs slightly between the co- and cross-polarized measurement, the bunching constants $c_{i}$ are set to zero within the evaluation, as to exclude this as an error for the calculation of the visibility. In this case, the normalized autocorrelation function is expected to drop to 0.5 for distinguishable photons and vanishing time delay. The measured value of $g^{(0)}_{\mathrm{HOM},\perp}(0)=0.463\pm0.097$ ($0.471\pm0.093$ before the deconvolution) is in good agreement with this. The autocorrelation for indistinguishable photons yields $g^{(2)}_{\mathrm{HOM},\parallel}(0)=0.049\pm0.04$ ($0.135\pm0.045$ before the deconvolution). The maximal degree of indistinguishability of the photons is calculated to $V_{\mathrm{HOM,decon}}=0.894\pm0.109$ ($V_{\mathrm{HOM,raw}}=0.713\pm0.15$) including (excluding) the deconvolution of the data with the system response function. The width of the central dip is a measure of the temporal post-selection window necessary for possible time-gated applications. The $1/e$ rise time is given by $T_{\mathrm{b}}$, from which a full width of $2T_{\mathrm{b}}=\SI{1.156\pm 0.137}{\nano \second}$ is determined. Apart from approaches relying on quantum frequency conversion~\cite{Weber2019}, this constitutes the first direct measurement of the mutual degree of indistinguishability of QD photons in the telecom C-band, complementing the demonstration of the three basic prerequisites for quantum applications, namely single-photon emission~\cite{Paul2017}, entangled-photon pair emission~\cite{Olbr} and indistinguishability.

In conclusion, a study on the coherence of InAs/InGaAs/GaAs QDs emitting in the telecom C-band was presented. Fourier-transform spectroscopy in AB pumping revealed a mean linewidth of \SI{9.74}{\giga \hertz} of transitions from 9 exemplary QDs due to a very strong influence of inhomogeneous broadening effects, motivating the change to resonant pumping schemes. In RF, the mean linewidth of five QDs is reduced to \SI{3.50}{\giga \hertz}. Furthermore, coherent state preparation with a fidelity of \SI{49.2}{\percent} in pulsed RF paves the way to on-demand generation of telecom C-band photons with a good coherence. Offering the inherent possibility of polarization-entangled photon pair emission, TPE is investigated as another resonant excitation scheme. The autocorrelation function of the XX line exhibits bunching behaviour on three different time scales as is typically observed in resonant pumping schemes. The single-photon purity yields a value of $g^{(2)}_{\mathrm{HBT,decon}}(0)=0.072\pm0.104$ ($g^{(2)}_{\mathrm{HBT,raw}}(0)=0.102\pm0.109$) including (excluding) a deconvolution with the system response function. Finally, the degree of indistinguishability of the XX transition is determined, yielding a value of $V_{\mathrm{HOM,decon}}=0.894\pm0.109$ ($V_{\mathrm{HOM,raw}}=0.713\pm0.15$). The presented results show that state-of-the-art values for quantum optical properties of InAs/GaAs QDs emitting in the near infrared are within reach for GaAs-based QDs emitting in the telecom C-band.

\begin{acknowledgments}
The authors gratefully acknowledge the funding by the German Federal Ministry of Education and Research (BMBF), in particular the projects Q.com-H (16KIS0115) and Q.Link.X (16KIS0862). The work reported in this paper was partially funded by project EMPIR 17FUN06 SIQUST. This project has received funding from the EMPIR programme co-financed by the Participating States and from the European Union’s Horizon 2020 research and innovation program. S. L. P. acknowledges the IQST center for the project QUDOSI. For fruitful discussions and assistance in the data evaluation, the authors wish to thank H. Vural, J. Maisch, M. Schwartz and F. Hornung. Furthermore, the reliable and persistent support by the company Scontel for their SSPD detector system is acknowledged. 
\end{acknowledgments}

\bibliographystyle{apsrev4-1}
\bibliography{2019_manuscript_nawrath_arxiv}

\end{document}